\documentclass[10pt]{article}
\usepackage{graphicx}
\usepackage{amsmath}
\usepackage{amssymb}
\usepackage{caption2}
\begin{document}
\begin{center}
{\large {\bf \sc{  Semi-leptonic  $B\to S$ decays in the standard model and  in the universal extra
dimension model }}} \\[2mm]
Zhi-Gang Wang \footnote{E-mail,zgwang@aliyun.com.  }    \\
 Department of Physics, North China Electric Power University,
Baoding 071003, P. R. China
\end{center}

\begin{abstract}
In this article,  we assume  the two nonets of scalar mesons below and above 1 GeV are all $\bar{q}q$ states, and
  study the semi-leptonic decays  $B\to S\ell^-\bar{\nu}_{\ell}$, $B\to S\ell^+\ell^-$ and $B\to S\bar{\nu}\nu$ both
  in the standard model and in the universal extra
dimension model using the  $B-S$ form-factors calculated by the light-cone QCD sum rules in our previous work.
 We obtain the partial decay widths and decay widths, which can be confronted with the experimental data in the future to examine
 the natures  of the scalar mesons and constrain the basic parameter in the universal extra
dimension model, the  compactification scale $1/R$.
\end{abstract}



\section{Introduction}

The natures of the scalar mesons are not well established theoretically, and their underlying structures  are under hot debating \cite{ReviewAmsler}.
Irrespective of the two-quark state, tetraquark state and glueball assignments, the  underlying structures determine their  productions and decays.
In previous work, we assume that the scalar mesons are all $\bar{q}q$ states,
  in case I, the scalar mesons $\{f_0(600)$, $a_0(980)$, $\kappa(800)$, $f_0(980) \}$ below 1 GeV are the ground states, in case II, the scalar mesons
    $\{f_0(1370)$, $a_0(1450)$, $K^*_0(1430)$, $f_0(1500) \}$
  above 1 GeV are the ground states; and study the $B-S$ transition form-factors   with the light-cone QCD sum rules \cite{Wang1409}.
 The transition form-factors in the semi-leptonic decays  are highly nonperturbative quantities.
  They   not only depend on the dynamics of strong interactions
among the quarks in the initial and final mesons, but also depend  on the under structures of the involved mesons.
In this article, we take the $B-S$ form-factors  as basic input parameters, and  study the semi-leptonic decays
$B\to S\ell^-\bar{\nu}_{\ell}$, $B\to S\ell^+\ell^-$ and $B\to S\bar{\nu}\nu$ both
in the standard model and  in the universal extra
dimension model  to examine the natures  of the scalar mesons and search for new physic beyond the standard model.

The semi-leptonic $B$-decays  are excellent subjects in studying  the CKM matrix elements
and CP violations in the standard model.
They also serve as  a powerful probe of new physics beyond the standard model in a complementary way
to the direct searches, the indirect probe plays an important role in identifying the new physics  and its properties \cite{B-EPJC}. At the quark level, the semi-leptonic $B\to S$ decays take place through the transitions
$b \to u(c) \ell^{-}\bar{\nu}_{\ell}$, $b \to s(d) \ell^{+}\ell^{-}$ and $b \to s(d) \bar{\nu}_{\ell}\nu_{\ell}$.
In the standard model, the decays $b \to u(c) \ell^{-}\bar{\nu}_{\ell}$ take place through the exchange of the intermediate $W$ boson at the tree-level,
while the decays $b \to s(d) \ell^{+}\ell^{-}$ and $b \to s(d) \bar{\nu}_{\ell}\nu_{\ell}$ take place through the penguin diagrams and other diagrams at the one-loop level. The processes induced by the flavor-changing neutral currents $b \to s(d)$
  provide the most sensitive and
 stringent  test of the standard model at the one-loop level.
The branching fractions of the semi-leptonic decays $\bar{B}^0(b\bar{d})\to S(u\bar{d})\ell^{-}\bar{\nu}_{\ell}$, $B^-(b\bar{u})\to S(u\bar{u})\ell^{-}\bar{\nu}_{\ell}$, $\bar{B}_s^0(b\bar{s})\to S(u\bar{s})\ell^{-}\bar{\nu}_{\ell}$  are expected to be large, which favors examining the theoretical predictions in the standard model. The branching fractions of the semi-leptonic decays $\bar{B}^0(b\bar{d})\to S(s\bar{d})\ell^+\ell^-$, $B^-(b\bar{u})\to S(s\bar{u})\ell^+\ell^-$, $\bar{B}^0_s(b\bar{s})\to S(s\bar{s})\ell^+\ell^-$, $\bar{B}^0(b\bar{d})\to S(s\bar{d})\bar{\nu}_{\ell}\nu_{\ell}$, $B^-(b\bar{u})\to S(s\bar{u})\bar{\nu}_{\ell}\nu_{\ell}$, $\bar{B}^0_s(b\bar{s})\to S(s\bar{s})\bar{\nu}_{\ell}\nu_{\ell}$ are expected to be small, which favors searching for new physics beyond the standard model.
New physics effects manifest themselves in the rare $B$-decays in two different ways,
either through new contributions to the Wilson coefficients or through new operators in
the effective Hamiltonians, which are absent in the standard model.

The universal extra
dimension (UED) models  are  promising  models among various models of the new physics beyond the standard model \cite{UED-PRT},
 where all standard model fields are allowed to propagate  in all available dimensions.
The simplest model  is the Appelquist, Cheng and Dobrescu
(ACD) model, which has only one extra universal dimension \cite{ACD2001}.
The topology of the fifth dimension is the orbifold $S^1/Z_2$,  the coordinate $y=x_5$ runs from
$0$ to $ 2\pi R$, where the $R$ is the compactification radius. The orbifold has two fixed points at $y = 0$ and
$y = \pi R$, respectively, the boundary conditions in the two fixed points determine the Kaluza-Klein (KK)
mode expansion of all the fields.
In four dimensions after compactification, there are the standard model fields, the series of
their KK partners and additional series of KK modes having no correspondence
to the standard model fields.  The only additional free
parameter  is the compactification scale $1/R$,   the masses of the new
KK particles and the interactions among KK particles and  the standard model particles are described
by the additional parameter $1/R$ and the parameters of the standard model.
The presence of the boundaries of the $S^1/Z_2$ orbifold breaks
translational invariance along the extra dimension and therefore leads to the violation of
the KK-number at the loop
level but still preserves  a $Z_2$ symmetry (or KK-parity). The KK-parity warrants the stability of the lightest KK-excitation and
provides a viable dark matter candidate,  and implies
  disappearance   of the tree-level KK contributions to the low energy processes occur  at the energy scales $\mu\ll 1/R$.
The local operators in the low energy effective Hamiltonians are the same
both in the standard model and in the ACD model, and the effects  of the KK modes  amount to  modifying
the Wilson coefficients \cite{Buras2003,Buras2004}.

 The ACD model has  potentially  many  phenomenological interest, such as the semi-leptonic and radiative $B$-decays \cite{Colangelo2006},
semi-leptonic $\Lambda_b$ decays \cite{Aliev2007}, etc.
The  electro-weak precision
tests  yield a bound of $1/R >  500 \rm{GeV}$  in case of a UED Higgs boson with the mass about 125 GeV \cite{EW-500,EW-500-H}. Other analysis of the  electro-weak
precise measurements and the inclusive
radiative $b \to s\gamma $ decay imply
$1/R >   600\, \rm{GeV}$ \cite{R600-1,R600-2}. While the LHC searches for the dilepton resonances lead to
$1/R \geq  715\, \rm{GeV}$ \cite{R715}.  In this article, we study the semi-leptonic $B\to S$ decays both
in the standard model and  in the UED model, and try to obtain some constraints on the parameter $1/R$.

The article is arranged as follows:  we derive the decay widths for
the semi-leptonic $B\to S$ transitions   in Sect.2;
in Sect.3, we present the numerical results and discussions; and Sect.4 is reserved for our
conclusions.

\section{The decay widths in the standard model and  in the universal extra dimension model   }
In the following, we write down  the effective
Hamiltonian ${\cal H}_{eff}$ responsible for the transitions $b \to u \ell^- \bar{\nu}_{\ell}$, $b \to s \ell^+ \ell^-$ and $b \to s \bar{\nu} \nu $  in
the standard model and in the UED model \cite{Buras2003,Buras2004,Misiak-NPB,Buras1995,BurasRMP},
\begin{eqnarray}
 {\cal H}_{eff} &=&\frac{G_F}{\sqrt{2}}V_{ub}\bar{ u}\gamma_{\alpha}(1-\gamma_5)b \, \bar{\ell}\gamma^{\alpha}(1-\gamma_5)\nu_{\ell}
 -\frac{G_F V_{tb}V_{ts}^*}{\sqrt{2}}\frac{e^2}{8\pi^2}\left\{ C_{9}^{eff}\bar{s}\gamma _{\alpha}(1-\gamma _{5})b\,\bar{\ell}\gamma ^{\alpha } \ell\right.
 \nonumber \\
 &&\left.+C_{10}^{eff}\bar{s}\gamma _{\alpha}
(1-\gamma _{5})b \,\bar{\ell}\gamma ^{\alpha }\gamma _{5} \ell  -\frac{2im_{b}C_7^{eff} }{q^{2}}\bar{s}\sigma _{\alpha \beta}(1+\gamma _{5})q^{\beta }b \,\bar{\ell}\gamma ^{\alpha }\ell \right\}  \nonumber\\
&&+\frac{G_F V_{tb}V_{ts}^*}{\sqrt{2}}\frac{e^2}{8\pi^2\sin^2\theta_W}\eta_X X(x_t)\bar{s}\gamma _{\alpha}(1-\gamma _{5})b\,\bar{\nu}_{\ell}\gamma ^{\alpha }(1-\gamma_5) \nu_{\ell} \, ,
 \end{eqnarray}
where we have neglected the terms proportional to $V_{ub}V_{us}^{*}$
according to the value $|V_{ub}V_{us}^{*}/V_{tb}V_{ts}^{*}|\sim 10^{-2}$.
No new operators are induced in the ACD model,  the effects of
the KK contributions are implemented by modifying the Wilson
coefficients which also depend on the additional parameter, the
compactification radius $R$.  In the present case, we only need to specify the revelent  Wilson
coefficients $C^{eff}_7$, $C^{eff}_9$, $C^{eff}_{10}$ and $X(x_t)$ \cite{Buras2003,Buras2004}.
In this article, we neglect the long-distance contributions come from
the four-quark operators near the $c\bar{c}$ resonances, such as the $J/\psi$, $\psi^\prime$, $\cdots$, which can be
experimentally removed by applying appropriate kinematical cuts in the
neighborhood of the resonances \cite{CC}.

Now, we write down the Wilson
coefficients $C^{eff}_7$, $C^{eff}_9$ and $C^{eff}_{10}$,  explicitly,
\begin{eqnarray}
C_{7}^{eff}\left(\mu,\frac{1}{R}\right)&=&\eta ^{\frac{16}{23}}C_{7}\left(\mu _{W},\frac{1}{R}\right)+\frac{8}{
3}\left(\eta ^{\frac{14}{23}}-\eta ^{\frac{16}{23}}\right)C_{8}\left(\mu _{W},\frac{1}{R}\right)+C_{2}\left(\mu _{W},\frac{1}{R}\right)\sum_{i=1}^{8}h_{i}\eta ^{a _{i}} \, , \nonumber\\
C_9^{eff}\left(\mu,\frac{1}{R}\right) & = & C_9^{NDR}\left(\mu,\frac{1}{R} \right)\eta(\hat{s}) + h(z, \hat{s})
\left( 3C_1 + C_2 + 3 C_3 + C_4 + 3C_5 + C_6 \right) \nonumber \\
&& - \frac{1}{2} h(1, \hat{s}) \left( 4 C_3 + 4 C_4 + 3C_5 + C_6 \right) \nonumber \\
& & - \frac{1}{2} h(0, \hat{s}) \left( C_3 + 3 C_4 \right)
+ \frac{2}{9} \left( 3 C_3 + C_4 + 3 C_5 +C_6 \right) \, ,\nonumber \\
C_{10}^{eff}\left(\mu,\frac{1}{R}\right)&=&-\frac{Y\left(x_{t},\frac{1}{R}\right)}{\sin ^{2}\theta _{W}} \, ,
\end{eqnarray}
where $\eta  =\frac{\alpha_s(\mu _{W})}{\alpha _{s}(\mu)}$,  $\hat{s}=\frac{q^2}{m_b^2}$, $z=\frac{m_c}{m_b}$,
\begin{eqnarray}
C_{2}\left(\mu _{W},\frac{1}{R}\right)&=&1\, , \nonumber\\
C_{7}\left(\mu _{W},\frac{1}{R}\right)&=&-\frac{1}{2}D^{\prime }\left(x_{t},\frac{1}{R}\right)\, , \nonumber\\
C_{8}\left(\mu _{W},\frac{1}{R}\right)&=&-\frac{1}{2}E^{\prime }\left(x_{t},\frac{1}{R}\right) \, ,\nonumber\\
C^{NDR}_{9}\left(\mu,\frac{1}{R} \right)&=&P_{0}^{NDR}+\frac{Y\left(x_{t},\frac{1}{R}\right)}{\sin ^{2}\theta _{W}}
-4Z\left(x_{t},\frac{1}{R}\right)+P_{E}E\left(x_{t}\right)\, ,
\end{eqnarray}
 \begin{eqnarray}
\eta(\hat{s}) & = & 1 + \frac{\alpha_s(\mu)}{\pi}\,\omega(\hat{s}) \, ,  \nonumber\\
\omega(\hat{s}) & = & - \frac{2}{9} \pi^2 - \frac{4}{3}\mbox{Li}_2(\hat{s}) - \frac{2}{3}
\log \hat{s} \log(1-\hat{s}) - \frac{5+4\hat{s}}{3(1+2\hat{s})}\log(1-\hat{s}) -\frac{2 \hat{s} (1+\hat{s}) (1-2\hat{s})}{3(1-\hat{s})^2
(1+2\hat{s})} \log \hat{s} \nonumber \\
& &   + \frac{5+9\hat{s}-6\hat{s}^2}{6 (1-\hat{s}) (1+2\hat{s})} \, ,  \nonumber\\
 h(z,\hat{s}) & = & -\frac{8}{9}\ln\frac{m_b}{\mu} - \frac{8}{9}\log z +
\frac{8}{27} + \frac{4}{9} x \nonumber\\
& & - \frac{2}{9} (2+x) |1-x|^{1/2} \left\{
\begin{array}{ll}
 \log\left| \frac{\sqrt{1-x} + 1}{\sqrt{1-x} - 1}\right| - i\pi, &
\mbox{for } x \equiv \frac{4z^2}{\hat s} < 1 \nonumber \\
2 \arctan \frac{1}{\sqrt{x-1}}, & \mbox{for } x \equiv \frac {4z^2}{\hat
s} > 1,
\end{array}
\right. \\
h(0, \hat{s})& = & \frac{8}{27} -\frac{8}{9} \log\frac{m_b}{\mu} - \frac{4}{9} \ln \hat{s}+ \frac{4}{9} i\pi \, .
 \end{eqnarray}

\begin{eqnarray}
P_0^{NDR}&=&\frac{\pi}{\alpha_s(m_W)}\left(-0.1875+\sum_{i=1}^8 p_i \eta^{a_i+1} \right)+1.2468+\sum_{i=1}^8\eta^{a_i}\left( r_i^{NDR}+s_i\eta\right)\, , \nonumber\\
P_E&=&0.1405+\sum_{i=1}^8q_i\eta^{a_i+1} \, , \nonumber\\
C_j&=&\sum_{i=1}^8 k_{ji}\eta^{a_i}\, , \,\,\, i=1-6\, ,
\end{eqnarray}
the $NDR$ is the abbreviation for naive dimensional regularization, and the numerical values of the  parameters $a_i$, $h_i$, $p_{i} $,
$r^{NDR}_{i}$, $s_{i}$, $q_{i}$, $k_{ji}$, $i=1-6$, $j=1-8$ are taken from Refs.\cite{Buras1995,BurasRMP}.

The Wilson coefficients  $F\left(x_t, \frac{1}{R}\right)$ generalize  the corresponding standard model Wilson coefficients $F_0(x_t)$ according to the formula,
\begin{eqnarray}
F\left(x_t, \frac{1}{R}\right) &=&F_0(x_t) + \sum_{n=1}^{\infty} F_n(x_t, x_n)\, ,
\end{eqnarray}
where $x_t={ m_t^2 \over m_W^2}$, $x_n={ m_n^2 \over m_W^2}$ and $m_n={n \over R}$. Now we write down the Wilson coefficients $X\left(x_{t},\frac{1}{R}\right)$,
$Y\left(x_{t},\frac{1}{R}\right)$, $Z\left(x_{t},\frac{1}{R}\right)$, $D^{\prime }\left(x_{t},\frac{1}{R}\right)$ and $E^{\prime }\left(x_{t},\frac{1}{R}\right)$,  explicitly \cite{Buras2003,Buras2004},
\begin{eqnarray}
X/Y/Z\left(x_{t},\frac{1}{R}\right) &=&X/Y/Z_{0}(x_{t})+\sum_{n=1}^{\infty}C_{n}(x_{t},x_{n}) \, ,\nonumber \\
D^{\prime }\left(x_{t},\frac{1}{R}\right)&=&D^{\prime }_0(x_{t})+\sum_{n=1}^{\infty}D^{\prime }_n(x_{t},x_{n})\, , \nonumber\\
E^{\prime }\left(x_{t},\frac{1}{R}\right)&=&E^{\prime }_0(x_{t})+\sum_{n=1}^{\infty}E^{\prime }_n(x_{t},x_{n})\, ,
\end{eqnarray}
where
\begin{eqnarray}
X_{0}(x_{t}) &=&\frac{x_{t}}{8}\left[\frac{x_{t}+2}{x_{t}-1}+\frac{3x_{t}-6}{(x_{t}-1)^{2}}\log x_{t} \right]\, , \nonumber\\
Y_{0}(x_{t}) &=&\frac{x_{t}}{8}\left[\frac{x_{t}-4}{x_{t}-1}+\frac{3x_{t}}{(x_{t}-1)^{2}}\log x_{t} \right]\, , \nonumber\\
Z_{0}(x_{t}) &=&\frac{18x_{t}^{4}-163x_{t}^{3}+259x_{t}^{2}-108x_{t}}{
144(x_{t}-1)^{3}} +\left[\frac{32x_{t}^{4}-38x_{t}^{3}-15x_{t}^{2}+18x_{t}}{72(x_{t}-1)^{4}}-\frac{1}{9}\right]\log x_{t}\, , \nonumber\\
D_{0}^{\prime }(x_{t})&=&\frac{x_t(8x_{t}^2+5x_{t} -7 )}{12(x_{t}-1)^{3}}+\frac{x_{t}^{2}(2-3x_{t})}{2(x_{t}-1)^{4}}\log x_{t}\, , \nonumber\\
E_{0}^{\prime }(x_{t})&=&\frac{x_{t}(x_{t}^{2}-5x_{t}-2)}{4(x_{t}-1)^{3}}+\frac{3x_{t}^{2}}{2(x_{t}-1)^{4}}\log x_{t}\, ,
\end{eqnarray}
the last term in $C_9^{NDR}\left( \mu, \frac{1}{R}\right)$ is numerically negligible, we take the approximation $E\left(\mu,\frac{1}{R} \right)=E_{0}(x_{t})$,
\begin{eqnarray}
E_{0}(x_{t})&=&\frac{x_{t}(x_{t}^{2}+11x_{t}-18)}{12(x_{t}-1)^{3}}+\frac{x_{t}^{2}(15-16x_t+4x_t^2)}{6(x_{t}-1)^{4}}\log x_{t}-\frac{2}{3}\log x_t\, .
\end{eqnarray}
The summation  of the coefficients $C_n(x_t, x_n)$, $D^\prime_n(x_t, x_n)$ and $E^\prime_n(x_t, x_n)$ over $n$ leads to the formula \cite{Buras2003,Buras2004},
\begin{eqnarray}
\sum_{n=1}^{\infty }C_{n}(x_{t},x_{n})&=&\frac{x_{t}(7-x_{t})}{16(x_{t}-1)}-
\frac{\pi m_{W}Rx_{t}}{16(x_{t}-1)^{2}}\left[3(1+x_{t})J\left(R,-\frac{1}{2}
\right)+(x_{t}-7)J\left(R,\frac{1}{2}\right)\right] \, , \nonumber\\
\sum_{n=1}^{\infty }D_{n}^{\prime }(x_{t},x_{n}) &=&\frac{x_{t}\left[37-x_{t}(44+17x_{t})\right]}{72(x_{t}-1)^{3}}
+\frac{\pi m_{W}R}{2}\left[\int_{0}^{1}dy\frac{2y^{\frac{1}{2}}+7y^{\frac{3}{2}
}+3y^{\frac{5}{2}}}{6} \coth (\pi m_{W}R\sqrt{y}) \right.  \nonumber \\
&&+\frac{x_{t}(3x_{t}-2)(1+3x_{t})}{6(x_{t}-1)^{4}}J\left(R,-\frac{1}{2}\right)
\nonumber \\
&&-\frac{x_{t}(1+3x_{t})+(2-3x_{t})\left[1-x_{t}(10-x_{t})\right]}{6(x_{t}-1)^{4}} J\left(R,\frac{1}{2}\right)
\nonumber\\
&&\left.+\frac{(3x_{t}-2)(3+x_{t})-\left[1-x_{t}(10-x_{t})\right]}{6(x_{t}-1)^{4}}J\left(R,\frac{3}{2}\right)
 -\frac{3+x_{t}}{6(x_{t}-1)^{4}}J\left(R,\frac{5}{2}\right) \right] \, ,\nonumber\\
\sum_{n=1}^{\infty }E_{n}^{\prime }(x_{t},x_{n}) &=&\frac{
x_{t}\left[17+x_{t}(8-x_{t})\right]}{24(x_{t}-1)^{3}}  +\frac{\pi m_{W}R}{4}\left[\int_{0}^{1}dy\left(y^{\frac{1}{2}}+2y^{\frac{3}{2}}-3y^{
\frac{5}{2}}\right)\coth (\pi m_{W}R\sqrt{y}) \right.  \nonumber \\
&&-\frac{x_{t}(1+3x_{t})}{(x_{t}-1)^{4}}J\left(R,-\frac{1}{2}\right)  +\frac{x_{t}(1+3x_{t})-\left[1+x_{t}(x_{t}-10)\right]}{(x_{t}-1)^{4}}
J\left(R,\frac{1}{
2}\right)  \nonumber \\
&&\left.-\frac{3+x_{t}-\left[1+x_{t}(x_{t}-10)\right]}{(x_{t}-1)^{4}}J\left(R,\frac{3}{2}\right)
 +\frac{3+x_{t}}{(x_{t}-1)^{4}}J\left(R,\frac{5}{2}\right)\right] \, ,
\end{eqnarray}
where
\begin{equation}
J(R,\alpha )=\int_{0}^{1}dyy^{\alpha }\left[\coth (\pi m_{W}R\sqrt{y})-x_{t}^{1+\alpha }\coth (\pi m_{t}R\sqrt{y})\right] \, .
\end{equation}
The masses of the  KK states increase monotonously with increase of the  value of $1/R$, in the limit $1/R \to \infty$, the KK states decouple from
the low-energy processes  and the standard model phenomenology are recovered.

The Wilson coefficients are calculated at the matching energy scale $\mu_W$, then they evolve to the energy scale $\mu$ with the re-normalization group equation,
$C(\mu)=U(\mu,\mu_W)C(\mu_W)$, the re-normalization group evolution matrix $U(\mu,\mu_W)$ is available at the next-to-leading order (NLO) \cite{Buras1995,BurasRMP} and partial next-to-next-to-leading order (NNLO) \cite{NNLO}, the initial condition $C(\mu_W)$ is available at the NNLO \cite{NNLO} in the standard model, where only the QCD corrections are taken into account.
If we switch on the ACD model, the initial condition $C_{ACD}(\mu_W)$ is only available  at the NLO \cite{Buras2003,Buras2004}, then evolves to the energy scale $\mu$, $C(\mu)=U_{QCD}(\mu,\mu_W)C_{ACD}(\mu_W)$, some unknown uncertainties are introduced as we should use the formula  $C(\mu)=U_{ACD}(\mu,\mu_W)C_{ACD}(\mu_W)$, re-normalization group evolution matrix $U_{ACD}(\mu,\mu_W)$ receives contributions from the ACD model. The re-normalization group evolution matrix $U_{ACD}(\mu,\mu_W)$ at the NLO is still unavailable, and the initial condition $C_{ACD}(\mu_W)$ at the NNLO is also unavailable. Although the mixing effects induced by the $U_{QCD}(\mu,\mu_W)C_{QCD}(\mu_W)$ at the NNLO are rather large, we should be cautious in switching on the ACD model just by hand. In this article, we take the approximation $C(\mu)=U_{QCD}(\mu,\mu_W)C_{ACD}(\mu_W)$
 at the NLO in studying the $b\to s\ell^+\ell^-$ decays.

Now we study the semi-leptonic decays  $B\to S\ell^- \nu_{\ell}$, $B\to S\ell^+\ell^-$, $B\to S\bar{\nu}_{\ell}\nu_{\ell}$ with the effective
Hamiltonian ${\cal H}_{eff}$ and write down the transition amplitudes,

\begin{eqnarray}
\langle \bar{\nu}_{\ell}(k_1)\ell^-(k_2)S(p)|{\cal H}_{eff} |B(p^\prime)\rangle &=&-\frac{G_F V_{ub}}{\sqrt{2}}\langle S(p)|\bar{u}(0)\gamma^{\alpha}\gamma_5b(0) |B(p^\prime)\rangle \bar{u}(k_2)\gamma_\alpha(1-\gamma_5) v(k_1)\, , \nonumber\\
\langle \ell^+(k_1)\ell^-(k_2)S(p)|{\cal H}_{eff} |B(p^\prime)\rangle &=&\frac{G_F V_{tb}V_{ts}^*}{\sqrt{2}}\frac{\alpha}{2\pi}\left\{C_7^{eff}\frac{2im_b}{q^2}\langle S(p)|\bar{s}(0)\sigma^{\alpha\beta}\gamma_5q_\beta b(0) |B(p^\prime)\rangle \right. \nonumber\\
&&\bar{u}(k_2)\gamma_\alpha v(k_1)+C_9^{eff}\langle S(p)|\bar{s}(0)\gamma^{\alpha}\gamma_5b(0) |B(p^\prime)\rangle \bar{u}(k_2)\gamma_\alpha v(k_1)\nonumber\\
&&\left.+C_{10}^{eff}\langle S(p)|\bar{s}(0)\gamma^{\alpha}\gamma_5b(0) |B(p^\prime)\rangle \bar{u}(k_2)\gamma_\alpha\gamma_5 v(k_1)\right\} \, , \nonumber\\
\langle \bar{\nu}(k_1)\nu(k_2)S(p)|{\cal H}_{eff} |B(p^\prime)\rangle &=&-\frac{G_F V_{tb}V_{ts}^*}{\sqrt{2}}\frac{\alpha}{2\pi\sin^2\theta_W}
\eta_X X(x_t)\langle S(p)|\bar{s}(0)\gamma^{\alpha}\gamma_5b(0) |B(p^\prime)\rangle \nonumber\\
&&\bar{u}(k_2)\gamma_\alpha(1-\gamma_5) v(k_1)  \, ,
\end{eqnarray}
 then we take into account the definitions for the transition form-factors,
\begin{eqnarray}
\langle S(p)|\overline{q}(0) \gamma_{\mu}\gamma_5 b(0)|B(p^\prime)\rangle&=&-2iF_{+}(q^2)p_\mu -i\left[ F_{+}(q^2)+F_{-}(q^2)\right]q_\mu\, , \nonumber\\
&=&-i\left[F_{1}(q^2)\left(P_\mu-\frac{m^2_{B}-m_S^2}{q^2}q_\mu\right)
+F_{0}(q^2)\frac{m^2_{B}-m_S^2}{q^2}q_\mu\right]\, ,\nonumber\\
\langle S(p)|\overline{q}(0) \sigma_{\mu\nu}\gamma_5q^\nu b(0)|B(p^\prime)\rangle&=&-\frac{2F_T(q^2)}{m_{B}+m_S}\left(q^2p_\mu- q\cdot p q_\mu \right)\, ,
\end{eqnarray}
where $P=p^\prime+p$, $p^\prime=p+q$, $q=k_1+k_2$ and
\begin{eqnarray}
F_{1}(q^2)&=&F_{+}(q^2)\, , \nonumber\\
F_{0}(q^2)&=&F_{+}(q^2)+\frac{q^2}{m^2_{B}-m_S^2}F_{-}(q^2)\, .
\end{eqnarray}
Finally we obtain the partial decay widths,

\begin{eqnarray}
\frac{d\Gamma(B\to S \ell^-\bar{\nu}_{\ell})}{dq^2}&=&\frac{G_F^2  |V_{ub}|^2}{384\pi^3m_B^3}\frac{\left(q^2-m_{\ell}^2\right)^2}{q^6}\sqrt{\lambda\left(m_B^2,m_S^2,q^2\right)}\nonumber\\
&& \left\{\lambda\left(m_B^2,m_S^2,q^2\right)\left(2q^2+m^2_{\ell}\right)F_1^2(q^2)
 +3m_{\ell}^2\left(m_B^2-m^2_S\right)^2F_0^2(q^2) \right\} \, ,
\end{eqnarray}

\begin{eqnarray}
\frac{d\Gamma(B\to S \ell^+\ell^-)}{dq^2}&=&\frac{G_F^2\alpha^2 |V_{tb}V_{ts}^*|^2}{512\pi^5m_B^3}\sqrt{\frac{q^2-4m_{\ell}^2}{q^2}}\frac{\sqrt{\lambda\left(m_B^2,m_S^2,q^2\right)}}{3q^2}\left\{ \left[ C_7^{eff*}\frac{2m_b F_T(q^2)}{m_B+m_S}+\right.\right.\nonumber\\
&&\left.C_9^{eff*}F_1(q^2)\right]\left[ C_7^{eff}\frac{2m_b F_T(q^2)}{m_B+m_S}+C_9^{eff}F_1(q^2)\right]\lambda\left(m_B^2,m_S^2,q^2\right)\nonumber\\
&&\left(q^2+2m^2_{\ell}\right)+C_{10}^{eff*}C_{10}^{eff}\left[ F_1^2(q^2)\lambda\left(m_B^2,m_S^2,q^2\right)\left(q^2-4m^2_{\ell}\right) \right.\nonumber\\
&&\left. \left.+F_0^2(q^2)6m_{\ell}^2\left(m_B^2-m^2_S\right)^2 \right]\right\} \, ,
\end{eqnarray}

\begin{eqnarray}
\frac{d\Gamma(B\to S \bar{\nu}\nu)}{dx}&=&\frac{3G_F^2\alpha^2 |V_{tb}V_{ts}^*|^2}{384\pi^5m_B\sin^4\theta_W}\eta_X^2 X^2(x_t)\lambda\left(m_B^2,m_S^2,q^2\right)
\sqrt{\lambda\left(m_B^2,m_S^2,q^2\right)} F_1^2(q^2) \, ,
\end{eqnarray}
where $x=\frac{E_{mis}}{m_B}=\frac{m_B^2+q^2-m_S^2}{2m_B^2}$, the $E_{min}$ denotes the missing energy in the decays  $B\to S \bar{\nu}\nu$ and $\lambda(a,b,c)=a^2+b^2+c^2-2ab-2bc-2ca$. In calculating the decay widths, it is more convenient to use the form-factors $F_1(q^2)$ and $F_0(q^2)$.

The decays  $B\to S\mu^+\mu^-$ receive both resonant  and non-resonant contributions,
the decays mediated by the favor-changing neutral currents  lead
to the $\mu^+\mu^-$ with a non-resonant mass distribution, while  the intermediate states $J/\psi$, $\psi^{\prime}$, $\cdots$  in the decays $B \to S J/\psi (\psi^{\prime})\to S\mu^+\mu^-$
lead to the  $\mu^+\mu^-$ with a resonant mass distribution. The LHCb collaboration observed  both the decay $B^+ \to \psi(4160)K^+$ and
the subsequent decay  $\psi(4160)\to \mu^+\mu^-$, the resonant decay and the interference
contribution make up $20\%$ of the yield for the $\mu^+\mu^-$ masses above $3770\,\rm{MeV}$ \cite{LHCb-BK}.
So it is important  that the intermediate $\bar{c}c$ loops which appear in the processes $b \to \bar{c}c s$ at the quark-level are properly accounted for.

At small hadronic recoil or at large $q^2$, we should take into account the resonant contributions consistently.  We can employ the QCD factorization
approach to account for both the  factorizable and non-factorizable hard-gluon contributions and calculate the  non-factorizable
soft-gluon contributions with the
light-cone QCD sum rules, then match those corrections to the hadronic dispersion relation with respect to
the variable $q^2$, and  continue $q^2$ to the kinematical region of the decay \cite{CC}.
However, the soft-gluon contributions  coming from the four-quark and penguin operators
induce a positive contribution to the $C_9^{eff}$, which enhances the anomaly in the decays $B\to K^* \mu^+\mu^-$ \cite{cc-NP}.
We can also take into account those effects with the operator product expansion \cite{OPE}, but the theoretical predictions are smaller than the experimental value \cite{LHCb-BK}.

In Ref.\cite{LZ-1406},  Lyon and Zwicky  study  the interference pattern of the charm-resonances $\psi$, $\psi^{\prime}$, $\psi(3770)$, 	$\psi(4040)$, 	$\psi(4160)$ and $\psi(4415)$ with the electro-weak penguin operator $\mathcal{O}_9$
  in the decays $B^+ \to K^+ \mu^+\mu^-$ in details.   The observed  interference pattern by the LHCb collaboration is opposite in sign
and significantly enhanced as compared to the factorization approximation. A change of the result based on factorization approximation by a factor of $-2.5$
leads to a reasonable agreement with the experimental data. The non-factorizable corrections  are color enhanced but $\frac{\alpha_s}{\pi}$-suppressed, the factor $-2.5$ seems odd.

In the present case, we cannot take into account the contributions from both the hand-gluon and soft-gluon emissions in the intermediate $\bar{c}c$ loops in a systematic way. It is difficult to take into account the soft-gluon contributions  as the three-particle light-cone distribution amplitudes of the scalar mesons are not well studied, even the natures of the scalar mesons are still under debate. On the other hand,
  the non-factorizable hand-gluon contributions are not studied yet, which are in contrast to the decays $B\to K(K^*)\mu^+ \mu^-$.
Experimentally, the branching fractions of the $B$-decays to $ J/\psi K^0$, $ J/\psi K^+$,   $J/\psi K^{*0}$ and $J/\psi K^{*+}$ are $(8.73\pm0.32) \times10^{-4}$, $(1.027\pm0.031) \times10^{-3}$, $(1.32\pm0.06) \times10^{-3}$ and $(1.44\pm0.08) \times10^{-3}$, respectively \cite{PDG}, the large branching fractions and small uncertainties  favor extracting the  transition amplitudes    $A_{B  J/\psi K}$ and $A_{B  J/\psi K^*}$ so as to take into account contributions of the intermediate $\bar{c}c$ resonances in the decays $B\to K(K^*)\mu^+ \mu^-$. For the scalar modes,  only the decays $B \to J/\psi f_0(600),\,J/\psi f_0(980),\,J/\psi a_0(980)$ are observed, the branching fractions are   $\left(6.5^{+2.6}_{-1.1}\right) \times10^{-6}$, $\leq 1.1 \times10^{-6}$, $(4.7\pm3.4) \times10^{-7}$, respectively
 \cite{PDG}, the tiny branching fractions therefore tiny transition amplitudes can be  neglected.  The decays $B \to J/\psi S$, $\psi^{\prime}S$, $\psi(4160)S$ revelent to the present processes are not observed yet, we cannot extract the transition amplitudes    $A_{B  J/\psi S}$, $A_{B\psi^{\prime}S}$, $A_{B\psi(4160)S}$ from the experimental data to   account for the contributions of the intermediate $\bar{c}c$ resonances as in Ref.\cite{CC}.
For the $B_s$ meson, only the decays $B_s \to J/\psi f_{0}(980),\,J/\psi f_0(1370)$ are observed.   In this article, we take into account  the  factorizable contributions   of the $\bar{c}c$ loops in the leading-order approximation, and neglect the contributions of the intermediate $\bar{c}c$ resonances according to the scarce experimental data on the decays  $B \to J/\psi S$, $\psi^{\prime}S$, $\psi(4160)S$.

\section{Numerical results and discussions}
The input parameters  are taken as
$m_{B^0}=5279.55\,\rm{MeV}$, $\tau_{B^0}=1.519\times 10^{-12}\,\rm{s}$,
$m_{B_s}=5366.7\,\rm{MeV}$, $\tau_{B_s}=1.512\times 10^{-12}\,\rm{s}$,
$m_{a_0(980)}=(980\pm20)\,\rm{MeV}$,
$m_{\kappa(800)}=(682\pm29)\,\rm{MeV}$,
$m_{f_0(980)}=(990\pm20)\,\rm{MeV}$,
$m_{a_0(1450)}=(1474\pm19)\,\rm{MeV}$,
$m_{K_0^*(1430)}=(1425\pm50)\,\rm{MeV}$,
$m_{f_0(1500)}=(1505\pm6)\,\rm{MeV}$,
$m_e=0.511\,\rm{MeV}$, $m_{\mu}=105.658\,\rm{MeV}$,
$m_\tau=1776.82\,\rm{MeV}$,
$\alpha=\frac{1}{137}$, $\sin^2\theta_W=0.23$, $G_F=1.16637\times 10^{-5}\,\rm{GeV}^{-2}$,
  $|V_{ub}|=0.00355 \pm 0.00015$, $|V_{ts}|=0.0405^{+0.0011}_{-0.0012}$, $|V_{tb}|=0.99914\pm 0.00005$ \cite{PDG},
$m_b=4.8\,\rm{GeV}$, $m_c=1.4\,\rm{GeV}$, $m_W=80\,\rm{GeV}$, $m_t=170\,\rm{GeV}$, $\Lambda_{\overline{MS}}=0.225$,
$\alpha_s(\mu)=\frac{4\pi}{\beta_0\log\left(\mu^2/\Lambda^2_{\overline{MS}}\right)}\left[ 1-\frac{\beta_1}{\beta_0^2}\frac{\log\log\left(\mu^2/\Lambda^2_{\overline{MS}}\right)}{\log\left(\mu^2/\Lambda^2_{\overline{MS}}\right)}\right]$,
$\beta_0=\frac{23}{3}$, $\beta_1=\frac{116}{3}$, $\mu=5.0\,\rm{GeV}$ and $\eta_X=1$
 \cite{Buras1995,BurasRMP}.

The CKM matrix element $|V_{ub}|=(4.41 \pm  0.15 {}^{+0.15}_{-0.19})\times 10^{-3}$ from the inclusive processes and $|V_{ub}|=(3.28 \pm 0.29) \times 10^{-3}$ from the exclusive processes, the difference is remarkable. In this article,
we take the values $|V_{ub}|=0.00355 \pm 0.00015$, $|V_{ts}|=0.0405^{+0.0011}_{-0.0012}$ and $|V_{tb}|=0.99914\pm 0.00005$ determined by using a global fit to
all available measurements and imposing the standard model constraints \cite{PDG}.
The values $|V_{ub}|=(3.28 \pm 0.29) \times 10^{-3}$ from the exclusive processes and  $|V_{ub}|=0.00355 \pm 0.00015$ from the global fit are compatible.
The uncertainties of the branching fractions ${\rm Br}(B\to S\ell \bar{\nu}_{\ell})$ originate from the uncertainty of the $V_{ub}$ are about $0.18\%$. The uncertainties of the branching fractions ${\rm Br}(B\to S\ell^+\ell^-,\,S\bar{\nu}\nu)$ originate from the uncertainties of the $V_{tb}$ and $V_{ts}$ are about $0.09\%$ and $0.00\%$ respectively. The uncertainties originate from the CKM matrix elements are very small and can be neglected.

In previous work \cite{Wang1409}, we   calculate the $B-S$ form-factors by taking into account the perturbative  ${\mathcal{O}}(\alpha_s)$
 corrections to the twist-2 terms using the light-cone   QCD sum rules, and fit the numerical values of the form-factors into the single-pole forms,
  \begin{eqnarray}
 F_{i}(q^2)&=&\frac{F_{i}(0)}{1-a_{i}\frac{q^2}{m_B^2}} \, ,
 \end{eqnarray}
 where $m_B=5.28\,\rm{GeV}$, $i=+,-,T$, the values of the $F_{i}(0)$ and $a_{i}$ are shown explicitly in Tables 1-3.
 In calculations, we observe that the uncertainties induced by the uncertainties $\delta a_i$ are greatly
amplified in the regions $(m_b-m_S)^2-2(m_b-m_S)\chi\leq q^2\leq(m_B-m_S)^2$ with $\chi\approx 500\,\rm{MeV}$, and even larger than the central values, while the uncertainties
originate from the uncertainties $\delta F_i(0)$ are moderate. In the light-cone QCD sum rules,
the operator product expansion  is valid at small and intermediate momentum transfer squared
$q^2$, $0\leq q^2 \leq(m_b-m_S)^2-2(m_b-m_S)\chi$, the extrapolations to large values of the $q^2$ is out of control.
So we only retain the uncertainties $\delta F_i(0)$ and neglect the uncertainties $\delta a_i$.

 \begin{table}
\begin{center}
\begin{tabular}{|c|c|c|c|c|c|c|c|c|}\hline\hline
                       & $F_{+}(0)$        & $a_{+}$                   \\ \hline
  $B-a_0(980) $        & $0.576\pm0.042$   & $0.987\pm0.251$             \\ \hline
  $B-\kappa(800)$      & $0.504\pm0.039$   & $0.988\pm0.266$                  \\ \hline
  $B_s-\kappa(800)$    & $0.442\pm0.033$   & $0.904\pm0.274$                  \\ \hline
  $B_s-f_0(980) $      & $0.448\pm0.032$   & $0.952\pm0.257$                 \\ \hline
  $B-a_0(1450) $       & $0.549\pm0.071$   & $0.743\pm0.656$                 \\ \hline
  $B-K_0^*(1430)$      & $0.523\pm0.070$   & $0.795\pm0.669$                    \\ \hline
  $B_s-K_0^*(1430)$    & $0.458\pm0.062$   & $0.885\pm0.644$                    \\ \hline
  $B_s-f_0(1500) $     & $0.470\pm0.059$   & $0.941\pm0.595$                  \\ \hline
         \hline
\end{tabular}
\end{center}
\caption{ The parameters of the  transition form-factors $F_{+}(q^2)$.    }
\end{table}

\begin{table}
\begin{center}
\begin{tabular}{|c|c|c|c|c|c|c|c|c|}\hline\hline
                           & $-F_{-}(0)$       & $a_{-}$          \\ \hline
  $B-a_0(980) $            & $0.414\pm0.036$   & $0.904\pm0.319$    \\ \hline
  $B-\kappa(800)$          & $0.390\pm0.034$   & $0.934\pm0.314$         \\ \hline
  $B_s-\kappa(800)$        & $0.340\pm0.030$   & $0.829\pm0.342$         \\ \hline
  $B_s-f_0(980) $          & $0.305\pm0.029$   & $0.830\pm0.377$        \\ \hline
  $B-a_0(1450) $           & $0.287\pm0.067$   & $0.190\pm1.445$        \\ \hline
  $B-K_0^*(1430)$          & $0.275\pm0.064$   & $0.330\pm1.402$           \\ \hline
  $B_s-K_0^*(1430)$        & $0.240\pm0.058$   & $0.518\pm1.353$           \\ \hline
  $B_s-f_0(1500) $         & $0.222\pm0.057$   & $0.565\pm1.418$         \\ \hline
         \hline
\end{tabular}
\end{center}
\caption{ The parameters of the  transition form-factors $F_{-}(q^2)$.    }
\end{table}

 \begin{table}
\begin{center}
\begin{tabular}{|c|c|c|c|c|c|c|}\hline\hline
                       & $F_{T}(0)$        & $a_{T}$                  \\ \hline
  $B-a_0(980) $        & $0.778\pm0.062$   & $0.961\pm0.278$            \\ \hline
  $B-\kappa(800)$      & $0.673\pm0.056$   & $0.970\pm0.288$                 \\ \hline
  $B_s-\kappa(800)$    & $0.596\pm0.049$   & $0.877\pm0.304$                 \\ \hline
  $B_s-f_0(980) $      & $0.596\pm0.048$   & $0.900\pm0.299$                \\ \hline
  $B-a_0(1450) $       & $0.693\pm0.112$   & $0.511\pm0.893$                \\ \hline
  $B-K_0^*(1430)$      & $0.657\pm0.109$   & $0.598\pm0.900$                   \\ \hline
  $B_s-K_0^*(1430)$    & $0.575\pm0.098$   & $0.718\pm0.874$                   \\ \hline
  $B_s-f_0(1500) $     & $0.570\pm0.095$   & $0.778\pm0.835$                 \\ \hline
         \hline
\end{tabular}
\end{center}
\caption{ The parameters of the   transition form-factors $F_{T}(q^2)$.    }
\end{table}

Now we study the semi-leptonic decays in the standard model firstly. In Figs.1-3, we plot the partial decay widths of the $B\to S \ell^-\bar{\nu}_{\ell}$, $B\to S\ell^+\ell^-$ and $B\to S\bar{\nu}\nu$ with variations of the squared momentum $q^2$ of the leptonic pairs and the fractions $x$ of the missing energies, respectively, which can be confronted with the experimental data in the future.   In Fig.2, there exist  small discontinuities in the decays to the final states $Se^+e^-$ and $S\mu^+\mu^-$, which originate from
 the discontinuities in the $h(z,\hat{s})$ and $h(1,\hat{s})$ functions,   the discontinuities disappear in the decays to the final states $S\tau^+\tau^-$, as the value
$q^2\geq 4m^2_{\tau}$ is large enough to warrant  that the variations of the $q^2$ do not pass the
discontinuities in the $h(z,\hat{s})$ and $h(1,\hat{s})$ functions. From the Figs.1-2, we can see that the branching fractions of the decays to the final states $S\ell^-\bar{\nu}_{\ell}$ and $S\ell^+\ell^-$ with $\ell=e,\mu$ are much larger than the ones of the corresponding final states $S\tau^-\bar{\nu}_{\tau}$ and $S\tau^+\tau^-$ due to the much larger available phase-space.

The numerical values of the total branching fractions are shown in Table 4. From the table, we can see that the branching fractions of the
decays induced by the transitions $b\to u \ell^- \bar{\nu}_{\ell}$, $b\to s \ell^+ \ell^-$ and $b\to s \bar{\nu} \nu$ are of the orders $10^{-4}$, $10^{-7}$ and $10^{-6}$, respectively. The magnitudes  are compatible with the ones from other works based on the light-cone QCD sum rules \cite{WYM08,WW10} and perturbative QCD \cite{Li09} and light-front quark model \cite{LFQM}.
The transitions $b\to u \ell^- \bar{\nu}_{\ell}$ take place at the tree-level through the intermediate $W$-boson, while the transitions  $b\to s \ell^+ \ell^-$ and $b\to s \bar{\nu} \nu$ take place at the loop-level, so the decays $B\to S \ell^- \bar{\nu}_{\ell}$ have the largest branching fractions.  Compared to the decays
$B\to S \bar{\nu} \nu$, the decays $B\to S \ell^+ \ell^-$ have even smaller branching fractions due to the smaller phase-space.   The semi-leptonic decays $B\to S \ell^-\bar{\nu}_{\ell}$ are optimal in testing the standard model predictions, we can examine the natures  of the scalar mesons by confronting the predictions to the experimental data in the future,  while the semi-leptonic $B\to S \ell^+ \ell^-$ are optimal in searching for new physics beyond standard model.

From Table 4, we can also see that the branching fractions of the decays to the light scalar mesons $a_0(980)$, $\kappa(800)$, $f_0(980) $ below 1 GeV are much larger than that of the corresponding decays to the heavy scalar mesons    $a_0(1450)$, $K^*_0(1430)$, $f_0(1500)$ above 1 GeV due to the much larger energy released in the decays.

We can take into account the finite widths of the scalar mesons in the decays $B\to S\ell^- \nu_{\ell}\to PP^{\prime}\ell^- \nu_{\ell}$, $B\to S\ell^+\ell^-\to PP^{\prime}\ell^+\ell^-$, $B\to S\bar{\nu}_{\ell}\nu_{\ell}\to PP^{\prime}\bar{\nu}_{\ell}\nu_{\ell}$ by the simple replacement,
\begin{eqnarray}
\delta(m^2_{PP^\prime}-m_S^2)&\to& \frac{m_S\Gamma_S(m^2_{PP^\prime})}{\pi}\frac{1}{(m^2_{PP^\prime}-m_S^2)^2+m_S^2\Gamma_S^2(m^2_{PP^\prime})}\, , \nonumber\\
\Gamma_S(m^2_{PP^\prime})&=&\Gamma_S \frac{m_S}{m_{PP^\prime}}\sqrt{\frac{m^2_{PP^\prime}-(m_P+m_{P^\prime})^2}{m^2_{S}-(m_P+m_{P^\prime})^2}} \, ,
\end{eqnarray}
as the scalar mesons are reconstructed according to the pseudoscalar meson pairs $PP^\prime$ ($\eta\pi,\,K\pi,\,\pi\pi$). The effects of the finite widths can be factorized out approximately, and amount to multiplying the branching fractions by the factor $f_S$,
\begin{eqnarray}
f_{S}&=&\int_{(m_P+m_{P^\prime})^2}^{(m_B-\delta)^2} dm^2_{PP^\prime}\frac{m_S\Gamma_S(m^2_{PP^\prime})}{\pi}\frac{1}{(m^2_{PP^\prime}-m_S^2)^2+m_S^2\Gamma_S^2(m^2_{PP^\prime})}\, ,
\end{eqnarray}
where the $\delta=m_{\ell},\,2m_{\ell},\,0$ for the decays $B\to S\ell^- \nu_{\ell}$, $B\to S\ell^+\ell^-$, $B\to S\bar{\nu}_{\ell}\nu_{\ell}$, respectively.
The resulting values of  the factors $f_{S}$ are shown explicitly in Table 4.
$\frac{\Gamma_{f_0(980)}}{m_{f_0(980)}}=7.1\%$,
$\frac{\Gamma_{a_0(980)}}{m_{a_0(980)}}=7.7\%$,
$\frac{\Gamma_{f_0(1500)}}{m_{f_0(1500)}}=7.2\%$,
$\frac{\Gamma_{a_0(1450)}}{m_{a_0(1450)}}=18.0\%$,
$\frac{\Gamma_{K^*_0(1430)}}{m_{K^*_0(1430)}}=18.9\%$,
 $\frac{\Gamma_{\kappa(800)}}{m_{\kappa(800)}}=80.2\%$ for the central values of the masses and widths. From Table 4, we can see that the factor $f_S$ decreases with the increase of the $\frac{\Gamma_S}{m_S}$. For the decays $B\to \kappa(800)\ell^- \nu_{\ell}$, $B\to \kappa(800)\ell^+\ell^-$, $B\to \kappa(800)\bar{\nu}_{\ell}\nu_{\ell}$, the factors $f_S\approx 0.65$, the effects of the finite width are large, and the corresponding branching fractions are greatly reduced. For other decays, the effects of the finite widths are mild, the line-shapes of the partial decay widths with variations of the $q^2$ are not distorted significantly,  comparing  to the experimental data in the futures  make sense.

In Figs.4-5, we plot the branching fractions of the semi-leptonic decays $B\to S \ell^+\ell^-$ and $B\to S \bar{\nu}\nu$ with variations of the compactification scale $1/R$, respectively. From the figures, we can see that the branching fractions decrease monotonously with increase of the values $1/R$, at the region $1/R\geq 800\,\rm{GeV}$, the branching fractions almost reach constants, i.e. the KK states almost decouple from the low energy observables, while at the region $1/R\leq 600\,\rm{GeV}$, the impact of the KK states on the decays $B\to S \ell^+\ell^-$ are significant, at the region $1/R\leq 400\,\rm{GeV}$, the impact of the KK states on the decays $B\to S \bar{\nu}\nu$ are significant. If the constraint $1/R \geq  715\, \rm{GeV}$ obtained from
 the LHC searches for dilepton resonances is robust \cite{R715}, the semi-leptonic decays $B\to S \ell^+\ell^-$ are not the optimal processes in studying the UED model.
In the limit $1/R\rightarrow \infty$ or $R\rightarrow 0$, the summation  of the coefficients $C_n(x_t, x_n)$, $D^\prime_n(x_t, x_n)$ and $E^\prime_n(x_t, x_n)$ over $n$ does not vanish, but approach some constants which are independent on the $R$. The constants modify the Wilson coefficients slightly, and lead to slightly larger branching fractions, it is  difficult to distinguish the new physics effects from the standard model contributions.
Experimentally, the semi-leptonic decays  $B\to S\ell^- \nu_{\ell}$, $B\to S\ell^+\ell^-$, $B\to S\bar{\nu}_{\ell}\nu_{\ell}$ have not been observed yet, precise measurements and more theoretical works are still needed to make quantitative conclusion, at the  present time, we can only obtain  qualitative conclusion about the impact of the KK states.

It is not optimistic to extract the impacts of the KK states in the semi-leptonic decays $B\to S \ell^+\ell^-$ in case of $1/R>600\,\rm{GeV}$, the tiny  new physics effects are buried in the uncertainties originate  from the $B-S$ form-factors. The uncertainties of the $B-S$ form-factors originate mainly  from the light-cone distribution   amplitudes of the scalar  mesons, which are not well known at the present time, there exist hot controversies about the structures of the scalar mesons.
In the case of $1/R>600\,\rm{GeV}$ ($400\,\rm{GeV}$), the  semi-leptonic decays $B\to S \ell^+\ell^-$ ($B\to S\bar{\nu}_{\ell}\nu_{\ell}$) are optimal channels to explore the natures of the scalar mesons in the standard model. On the other hand, the semi-leptonic decays $B\to S\ell^- \nu_{\ell}$ take place at the tree level in the standard model, and have much larger branching fractions and also favor testing the natures of the scalar mesons, we can confronted the present predictions to the experimental data in the future.

\begin{figure}
 \centering
  \includegraphics[totalheight=5cm,width=7cm]{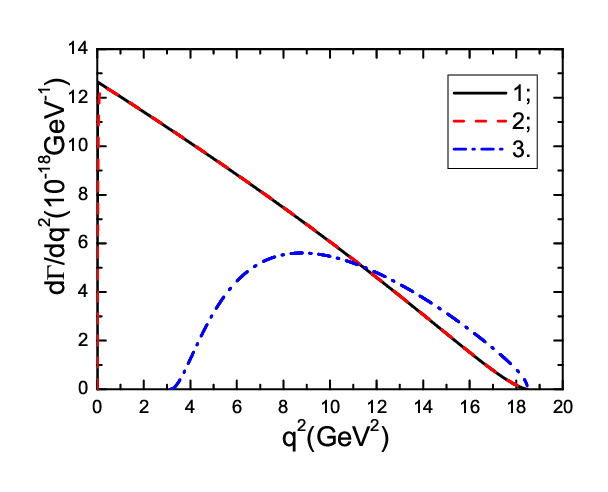}
  \includegraphics[totalheight=5cm,width=7cm]{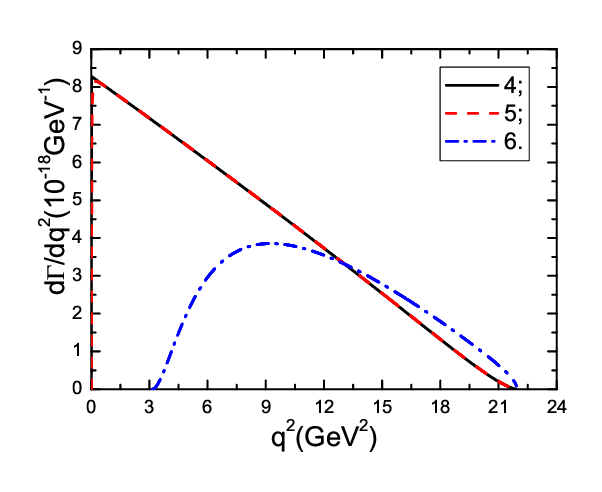}
  \includegraphics[totalheight=5cm,width=7cm]{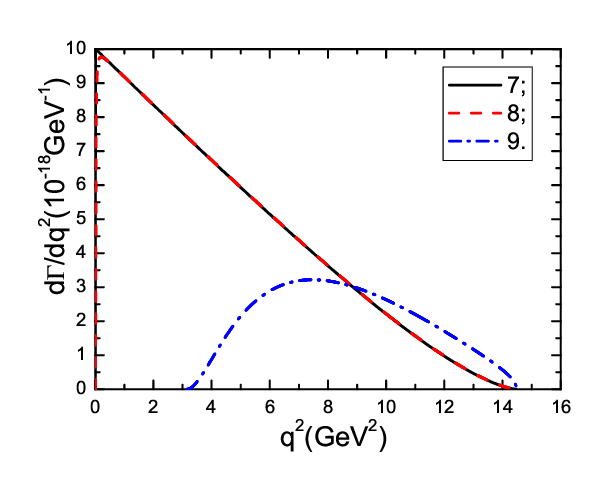}
  \includegraphics[totalheight=5cm,width=7cm]{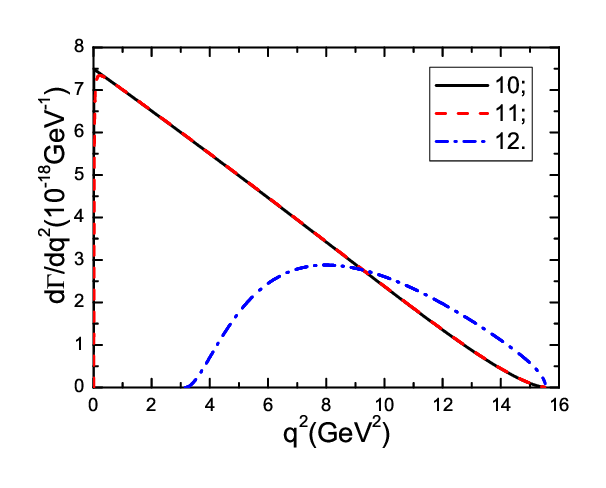}
    \caption{The partial decay widths  with variations of the   $q^2$, where the 1, 2, 3, 4, 5, 6, 7, 8, 9, 10, 11 and 12 denote the semi-leptonic decays
    $\bar{B}^0\to a^+_0(980)e^-\bar{\nu}_e $, $\bar{B}^0\to a^+_0(980)\mu^-\bar{\nu}_\mu $, $\bar{B}^0\to a^+_0(980)\tau^-\bar{\nu}_\tau $,
$\bar{B}_s\to \kappa^+(800)e^-\bar{\nu}_e $, $\bar{B}_s\to \kappa^+(800)\mu^-\bar{\nu}_\mu $, $\bar{B}_s\to \kappa^+(800)\tau^-\bar{\nu}_\tau $,
$\bar{B}^0\to a^+_0(1450)e^-\bar{\nu}_e $, $\bar{B}^0\to a^+_0(1450)\mu^-\bar{\nu}_\mu $, $\bar{B}^0\to a^+_0(1450)\tau^-\bar{\nu}_\tau $,
$\bar{B}_s\to K_0^{*+}(1430)e^-\bar{\nu}_e $,
$\bar{B}_s\to K_0^{*+}(1430)\mu^-\bar{\nu}_\mu $ and
$\bar{B}_s\to K_0^{*+}(1430)\tau^-\bar{\nu}_\tau $,  respectively.}
\end{figure}

\begin{figure}
 \centering
  \includegraphics[totalheight=5cm,width=7cm]{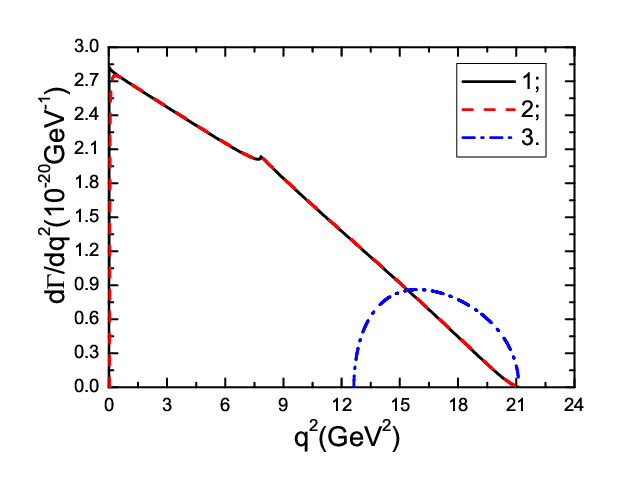}
  \includegraphics[totalheight=5cm,width=7cm]{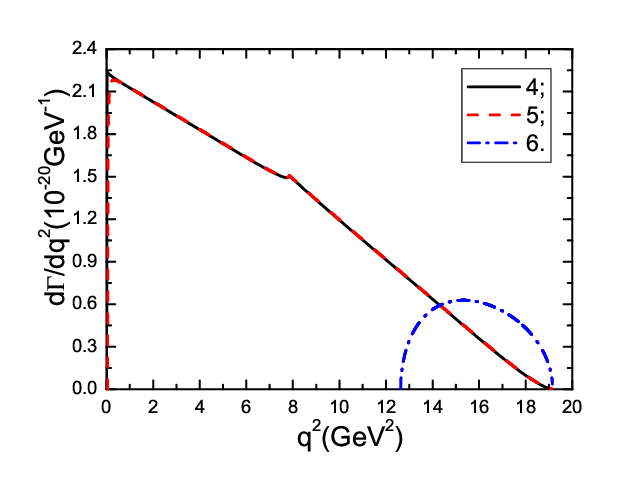}
  \includegraphics[totalheight=5cm,width=7cm]{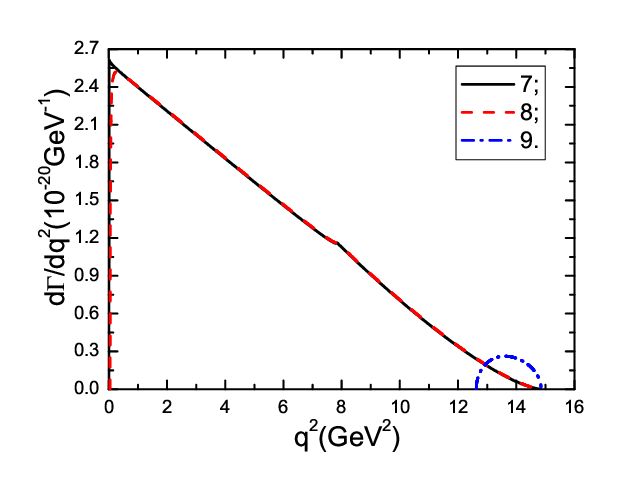}
  \includegraphics[totalheight=5cm,width=7cm]{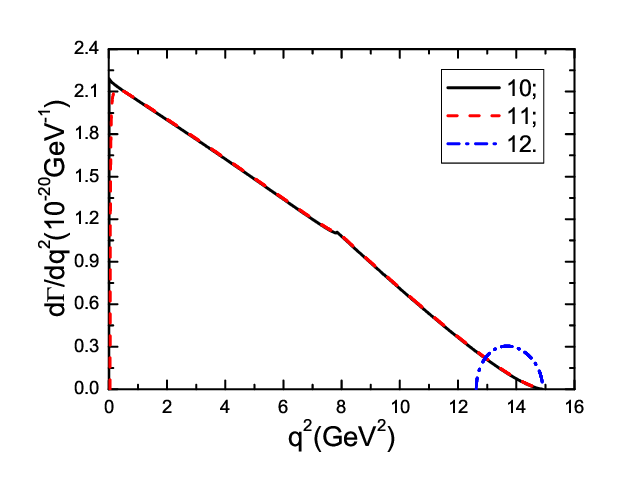}
    \caption{The partial decay widths  with variations of the   $q^2$, where the 1, 2, 3, 4, 5, 6, 7, 8, 9, 10, 11 and 12 denote the semi-leptonic decays
     $\bar{B}^0\to \kappa^0(800)e^+e^- $,  $\bar{B}^0\to \kappa^0(800)\mu^+\mu^- $,
$\bar{B}^0\to \kappa^0(800)\tau^+\tau^- $,  $\bar{B}_s\to f_0(980)e^+e^- $,
 $\bar{B}_s\to f_0(980)\mu^+\mu^- $, $\bar{B}_s\to f_0(980)\tau^+\tau^- $,
 $\bar{B}^0\to K^{*0}_0(1430)e^+e^- $,  $\bar{B}^0\to K^{*0}_0(1430)\mu^+\mu^- $,
$\bar{B}^0\to K^{*0}_0(1430)\tau^+\tau^- $,  $\bar{B}_s\to f_0(1500)e^+e^- $,
 $\bar{B}_s\to f_0(1500)\mu^+\mu^- $ and
$\bar{B}_s\to f_0(1500)\tau^+\tau^- $,  respectively.}
\end{figure}

\begin{figure}
 \centering
  \includegraphics[totalheight=6cm,width=8cm]{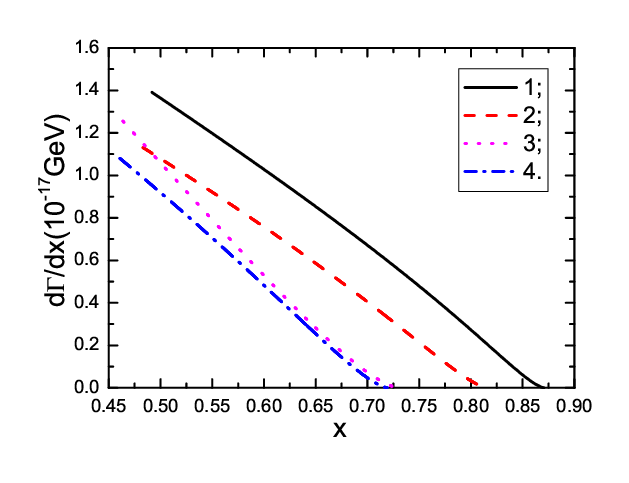}
    \caption{The partial decay widths  with variations of the   $x$, where the 1, 2, 3 and 4 denote the semi-leptonic decays
    $\bar{B}^0\to \kappa^0(800)\bar{\nu}\nu$,
$\bar{B}_s\to f_0(980)\bar{\nu}\nu $,
$\bar{B}^0\to K^{*0}_0(1430)\bar{\nu}\nu $ and
$\bar{B}_s\to f_0(1500)\bar{\nu}\nu $,  respectively.}
\end{figure}

 \begin{table}
\begin{center}
\begin{tabular}{|c|c|c|c|c|c|c|c|c|}\hline\hline
                          Decay channels               & Branching fractions                     & $f_S$\\ \hline
  $\bar{B}^0\to a^+_0(980)e^-\bar{\nu}_e $             & $(2.74\pm0.40\pm0.04)\times 10^{-4}$    & $0.95$ \\
  $\bar{B}^0\to a^+_0(980)\mu^-\bar{\nu}_\mu $         & $(2.74\pm0.40\pm0.04)\times 10^{-4}$    & $0.95$        \\
  $\bar{B}^0\to a^+_0(980)\tau^-\bar{\nu}_\tau $       & $(1.31\pm0.23\pm0.03)\times 10^{-4}$    & $0.95$  \\      \hline

  $\bar{B}_s\to \kappa^+(800)e^-\bar{\nu}_e $          & $(2.06\pm0.31\pm0.03)\times 10^{-4}$    & $0.65$ \\
  $\bar{B}_s\to \kappa^+(800)\mu^-\bar{\nu}_\mu $      & $(2.06\pm0.31\pm0.03)\times 10^{-4}$    & $0.65$ \\
  $\bar{B}_s\to \kappa^+(800)\tau^-\bar{\nu}_\tau $    & $(1.07\pm0.19\pm0.02)\times 10^{-4}$    & $0.63$ \\      \hline

  $\bar{B}^0\to a^+_0(1450)e^-\bar{\nu}_e $            & $(1.48\pm0.38\pm0.02)\times 10^{-4}$    & $0.92$ \\
  $\bar{B}^0\to a^+_0(1450)\mu^-\bar{\nu}_\mu $        & $(1.47\pm0.38\pm0.02)\times 10^{-4}$    & $0.92$ \\
  $\bar{B}^0\to a^+_0(1450)\tau^-\bar{\nu}_\tau $      & $(0.54\pm0.15\pm0.02)\times 10^{-4}$    & $0.91$ \\      \hline

  $\bar{B}_s\to K_0^{*+}(1430)e^-\bar{\nu}_e $         & $(1.27\pm0.35\pm0.06)\times 10^{-4}$    & $0.91$ \\
  $\bar{B}_s\to K_0^{*+}(1430)\mu^-\bar{\nu}_\mu $     & $(1.27\pm0.35\pm0.06)\times 10^{-4}$    & $0.91$ \\
  $\bar{B}_s\to K_0^{*+}(1430)\tau^-\bar{\nu}_\tau $   & $(0.54\pm0.16\pm0.04)\times 10^{-4}$    & $0.91$ \\      \hline

  $\bar{B}^0\to \kappa^0(800)e^+e^- $                  & $(7.34\pm1.22\pm0.13)\times 10^{-7}$    & $0.65$ \\
  $\bar{B}^0\to \kappa^0(800)\mu^+\mu^- $              & $(7.31\pm1.21\pm0.13)\times 10^{-7}$    & $0.65$ \\
 $\bar{B}^0\to \kappa^0(800)\tau^+\tau^- $             & $(1.33\pm0.36\pm0.09)\times 10^{-7}$    & $0.54$\\      \hline

  $\bar{B}_s\to f_0(980)e^+e^- $                       & $(5.16\pm0.79\pm0.07)\times 10^{-7}$    & $0.97$ \\
  $\bar{B}_s\to f_0(980)\mu^+\mu^- $                   & $(5.14\pm0.78\pm0.07)\times 10^{-7}$    & $0.97$ \\
 $\bar{B}_s\to f_0(980)\tau^+\tau^- $                  & $(0.74\pm0.17\pm0.04)\times 10^{-7}$    & $0.96$\\      \hline

  $\bar{B}^0\to K^{*0}_0(1430)e^+e^- $                 & $(4.14\pm1.17\pm0.18)\times 10^{-7}$    & $0.91$ \\
  $\bar{B}^0\to K^{*0}_0(1430)\mu^+\mu^- $             & $(4.12\pm1.17\pm0.18)\times 10^{-7}$    & $0.91$ \\
 $\bar{B}^0\to K^{*0}_0(1430)\tau^+\tau^- $            & $(0.11\pm0.03\pm0.04)\times 10^{-7}$    & $0.79$\\      \hline

  $\bar{B}_s\to f_0(1500)e^+e^- $                      & $(3.74\pm0.99\pm0.02)\times 10^{-7}$    & $0.97$ \\
  $\bar{B}_s\to f_0(1500)\mu^+\mu^- $                  & $(3.72\pm0.99\pm0.02)\times 10^{-7}$    & $0.97$ \\
 $\bar{B}_s\to f_0(1500)\tau^+\tau^- $                 & $(0.13\pm0.04\pm0.00)\times 10^{-7}$    & $0.92$\\      \hline

  $\bar{B}^0\to \kappa^0(800)\bar{\nu}\nu$             & $(6.30\pm0.97\pm0.11)\times 10^{-6}$    & $0.65$ \\
  $\bar{B}_s\to f_0(980)\bar{\nu}\nu $                 & $(4.39\pm0.63\pm0.06)\times 10^{-6}$    & $0.97$ \\
  $\bar{B}^0\to K^{*0}_0(1430)\bar{\nu}\nu $           & $(3.49\pm0.93\pm0.15)\times 10^{-6}$    & $0.91$ \\
  $\bar{B}_s\to f_0(1500)\bar{\nu}\nu $                & $(3.12\pm0.78\pm0.02)\times 10^{-6}$    & $0.97$ \\
        \hline
         \hline
\end{tabular}
\end{center}
\caption{ The branching fractions in the standard model and the finite width induced factors $f_S$, where the uncertainties come from the form-factors $B-S$ and masses $m_S$, respectively.    }
\end{table}

\begin{figure}
 \centering
  \includegraphics[totalheight=5cm,width=7cm]{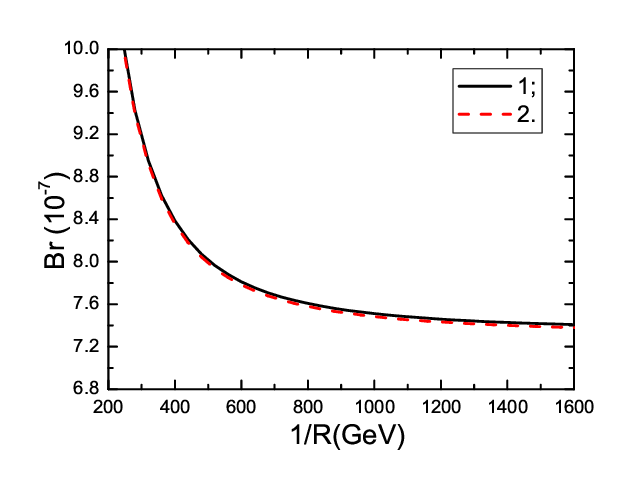}
  \includegraphics[totalheight=5cm,width=7cm]{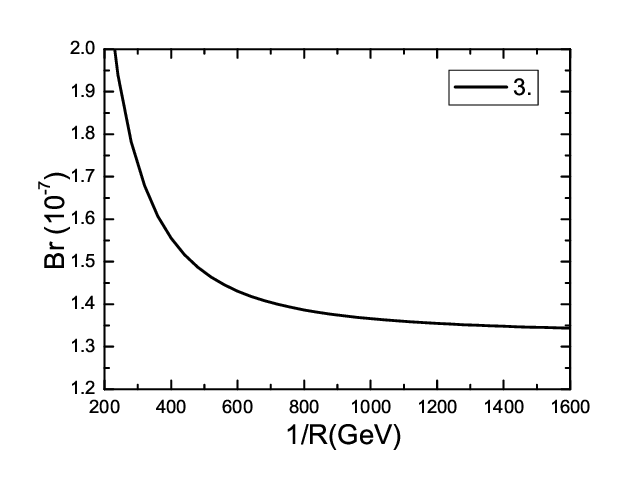}
  \includegraphics[totalheight=5cm,width=7cm]{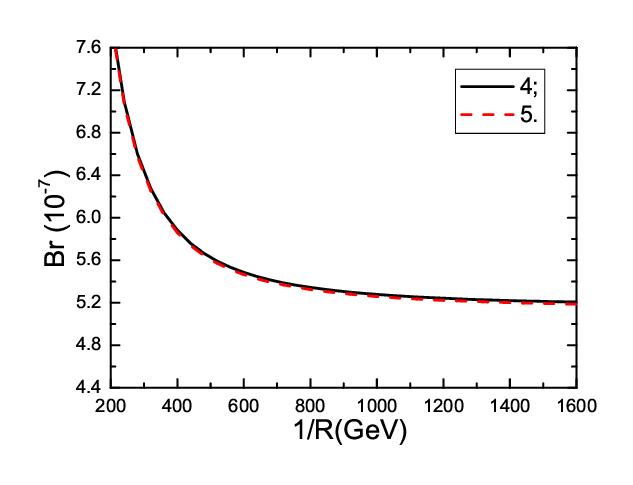}
  \includegraphics[totalheight=5cm,width=7cm]{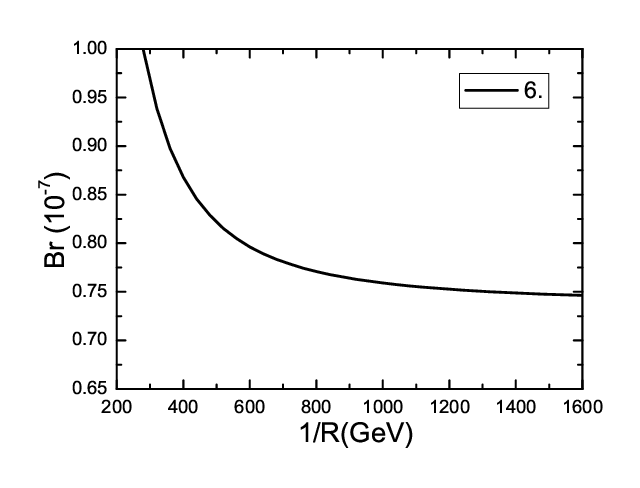}
  \includegraphics[totalheight=5cm,width=7cm]{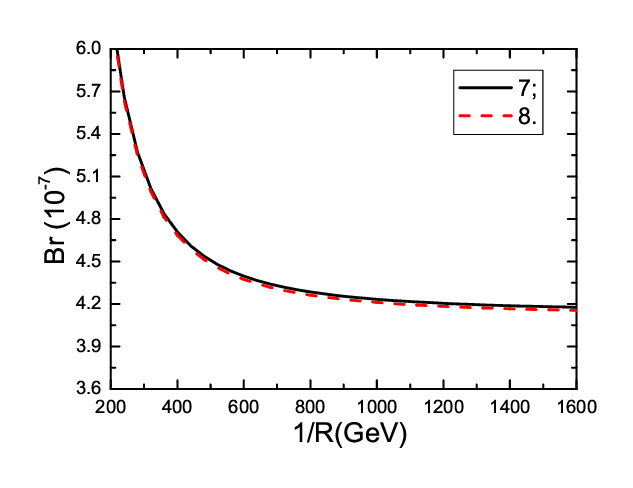}
   \includegraphics[totalheight=5cm,width=7cm]{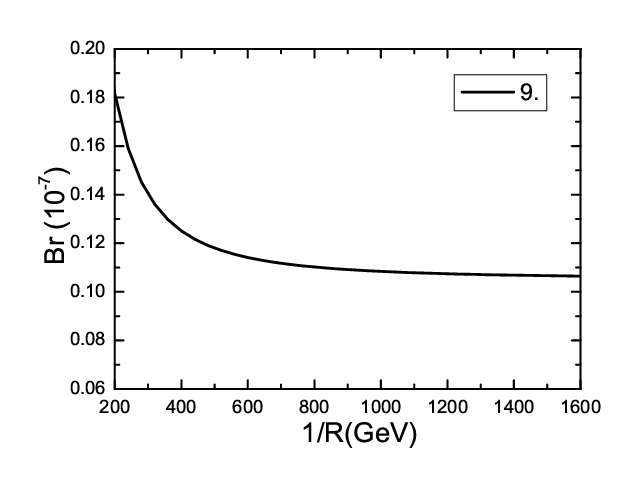}
  \includegraphics[totalheight=5cm,width=7cm]{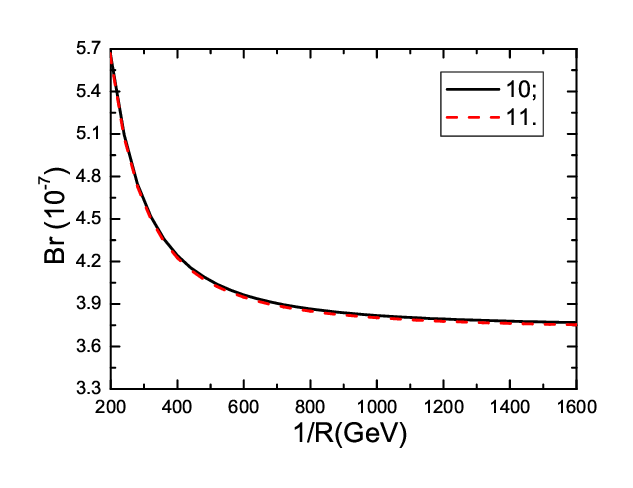}
  \includegraphics[totalheight=5cm,width=7cm]{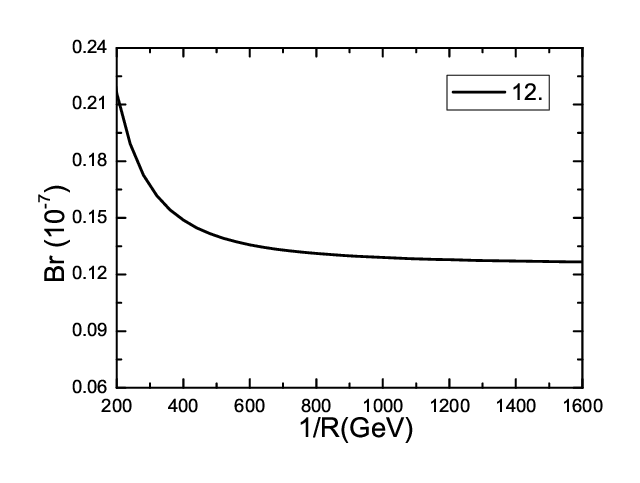}
    \caption{The branching fractions  with variations of the   $1/R$, where the 1, 2, 3, 4, 5, 6, 7, 8, 9, 10, 11 and 12 denote the semi-leptonic decays
     $\bar{B}^0\to \kappa^0(800)e^+e^- $,  $\bar{B}^0\to \kappa^0(800)\mu^+\mu^- $,
$\bar{B}^0\to \kappa^0(800)\tau^+\tau^- $,  $\bar{B}_s\to f_0(980)e^+e^- $,
 $\bar{B}_s\to f_0(980)\mu^+\mu^- $, $\bar{B}_s\to f_0(980)\tau^+\tau^- $,
 $\bar{B}^0\to K^{*0}_0(1430)e^+e^- $,  $\bar{B}^0\to K^{*0}_0(1430)\mu^+\mu^- $,
$\bar{B}^0\to K^{*0}_0(1430)\tau^+\tau^- $,  $\bar{B}_s\to f_0(1500)e^+e^- $,
 $\bar{B}_s\to f_0(1500)\mu^+\mu^- $ and
$\bar{B}_s\to f_0(1500)\tau^+\tau^- $,  respectively.}
\end{figure}

\begin{figure}
 \centering
  \includegraphics[totalheight=6cm,width=8cm]{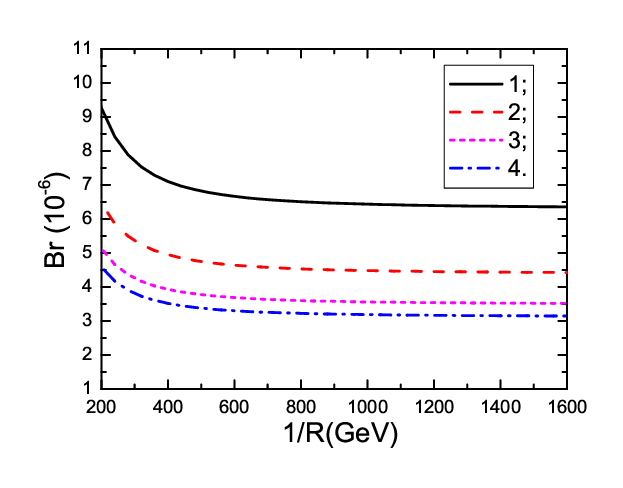}
    \caption{The branching fractions  with variations of the   $1/R$, where the 1, 2, 3 and 4 denote the semi-leptonic decays
    $\bar{B}^0\to \kappa^0(800)\bar{\nu}\nu$,
$\bar{B}_s\to f_0(980)\bar{\nu}\nu $,
$\bar{B}^0\to K^{*0}_0(1430)\bar{\nu}\nu $ and
$\bar{B}_s\to f_0(1500)\bar{\nu}\nu $,  respectively.}
\end{figure}

\section{Conclusion}
In previous work, we assume  the two nonets of scalar mesons below and above 1 GeV are all $\bar{q}q$ states,
  in case I, the scalar mesons below 1 GeV are the ground states, in case II, the scalar mesons above 1 GeV are the ground states, and
    calculate the $B-S$ form-factors by taking into account the perturbative  ${\mathcal{O}}(\alpha_s)$
 corrections to the twist-2 terms using the light-cone   QCD sum rules. In this article, we take those form-factor as basic input parameters, and
    study the semi-leptonic decays  $B\to S\ell^-\bar{\nu}_{\ell}$, $B\to S\ell^+\ell^-$ and $B\to S\bar{\nu}\nu$ both in the standard model and in the UED model.
 We obtain the partial decay widths and decay widths, which can be confronted with the experimental data in the future to examine
 the natures  of the scalar mesons and constrain the basic parameters in the UED model, the  compactification scale $1/R$.

\section*{Acknowledgements}
This  work is supported by National Natural Science Foundation,
Grant Numbers 11375063, and Natural Science Foundation of Hebei province, Grant Number A2014502017.

\end{document}